# How should the contact angle of a noncircular wetting boundary be described?

Jianhui Zhang*,#, Xiaosheng Chen#, Zhenzhen Gui*, Zhenlin Chen, Mingdong Ma, Yuxuan Huo, Weirong Zhang, Fan Zhang, Xiaosi Zhou, Xi Huang

School of Mechanical and Electrical Engineering, Guangzhou University, 230 Wai Huan Xi Road, Guangzhou Higher Education Mega Center, Guangzhou 510006, China

*Corresponding author. E-mail address: zhangjh@nuaa.edu.cn (Jianhui Zhang)

zhenzhengui@gzhu.edu.cn (Zhenzhen Gui)

#: These authors contributed equally

Tel.: +86 (20) 39366932 Fax: +86 (20) 39366932

## Abstract

For over 200 years, wettability has made significant contributions to understanding the properties of objects, advancing technological progress[1]. Theoretical model of the contact angle[2] (CA) for evaluating wettability has constantly been modified to address relevant emerging issues. However, these existing models[3-7] disregard the difference in the CA along the contact line and use a single-point CA to evaluate the entire contact line. From this perspective, there is no reasonable explanation for noncircular wetting. Here, we reveal that noncircular wetting boundaries result from property differences in the surfaces along the boundary, and utilize friction as a comprehensive factor reflecting local wettability. Average CA is proposed to evaluate the contact line instead of the single-point CA, making the Cassie method and Wenzel method obsolete, which will take an average property of the whole surface as a weight coefficient of the single-point CA, ignoring the subordination between physical properties and roughness in systematics.

**Main**

Historically, the Young-Dupré equation has been modified in one of three ways: (a) from the perspective of physics, considering the different surface energies between various materials at heterogeneous surfaces (Cassie equation), (b) in terms of engineering, considering the morphology and roughness of surfaces (Wenzel equation), and (c) simultaneously considering the above two perspectives. The core ideology guiding the modification of the Young-Dupré equation is to consider influencing factors and modes of an actual solid plane (ASP). The effects of surface energy and roughness on the droplet are averaged as correction coefficients to offset the deviation of the contact angle (CA) between the ASP and ideal solid plane (ISP). However, to reflect the properties of an ASP, only a single-point CA is measured. The properties of a single point are unable to represent the properties of a surface, which likewise cannot be equivalent to the properties of a point. Due to nonequivalent substitution, there are contradictions between theoretical models and actual measurements. In addition, in a scientific classification, the concept of roughness should be described by that of surface energy. Surface energy is a concept in physics, and roughness is a concept in engineering, which is extended from physics. However, wettability has always been discussed from the perspective of surface energy and roughness separately, ignoring the effects of interaction.

In the past, researchers mainly focused on the macroproperties of surfaces, and microscopic differences in the distribution of surface properties were ignored. A one-sided understanding of wettability did not cause severe consequences during the slow stage of the advancement of physics and material science. With advances in materials and preparation technology, various micro/nanochimaeras and complex structures have emerged, causing an increasing number of problems in both theoretical research and practical applications[8-14]. Differences in microscopic properties seem insignificant in the study of macrostructures but are indispensable in the study of micro/nanostructures and thus should no longer be ignored, as a slight difference in microscopic properties can cause a qualitative change in micro/nanostructure behaviour. On an ASP, metal droplets form a complex and irregular wetting boundary[15-22] after cooling. In further observation of other liquid droplets (such as water) on an ASP, noncircular wetting boundaries were ubiquitous[13, 23-30]. Analogously, Carmeliet et al. found that the local CA varied along the noncircular wetting boundary on a heterogeneous surface[26]. Regrettably, this was not explained with effective methods and was eventually evaluated through Cassie's apparent CA. Herein, we explore the unknown CA in noncircular wetting.

**Noncircular wetting boundary**

It is assumed that there is a kind of material particle with an absolutely uniform particle size and stable physical properties corresponding to the comprehensive physical properties of different materials. These particles can be arranged in accordance with certain rules to form a solid plane with isotropic surface tension and uniform surface roughness. This plane is called the ISP, corresponding to the actual heterogeneous solid plane (AHSP) composed of various materials. The equivalence of the ISP and AHSP is reflected in the fact that the entire surface energy and roughness of the solid plane are invariable and that the macroscopic surface tension $\gamma$ is consistent. When an ideal droplet spreads from the critical state of contact with the solid plane, part of the free energy of the gas-liquid interface is converted into that of the solid-liquid interface, and work is done over the rough plane. Due to the equivalent roughness between the ISP and AHSP, the friction does equal work. Accordingly, under identical conditions, the free energy of the liquid-solid interface formed by droplet spreading on the ISP is equal to that on the AHSP. Additionally, from the interface free energy equation $W=\gamma\delta A$, it can be inferred that the droplet spread areas on the ISP and AHSP are equal.

On the ISP, the wetting boundary formed by an ideal droplet is expected to be an ideal circle owing to the isotropic surface properties. Of course, this should also be realized without external interferences, such as pressure and temperature. However, the wetting boundary on the AHSP takes the form of a noncircular closed curve on account of the anisotropic surface properties. Furthermore, this boundary is expected to have an irregular geometry with concave and convex shapes around the ideal circle and the same area as the ideal circle. This is called the noncircular wetting boundary, which is a common result of differences in surface energy and roughness on the contact line. The paraffin-tin heterogeneous plane (PT plane) is taken as an example. Paraffin is a hydrophobic material, and tin is a hydrophilic material. The Young-Dupré CA of pure paraffin is 105°, while that of pure tin is approximately 75°. Hence, the CA of the droplet on the PT plane should be between 75° and 105°. Since hydrophilic and hydrophobic materials coexist, the surface properties vary at different locations over the PT plane. When a droplet spreads out on the PT plane, it presents a noncircular wetting boundary (Figure 1c-d). This phenomenon is common in actual surface wetting and becomes more prominent at smaller scales. In general, this phenomenon can be identified at the microscopic scale, but the wetting boundary looks like a circle at the macroscopic scale.

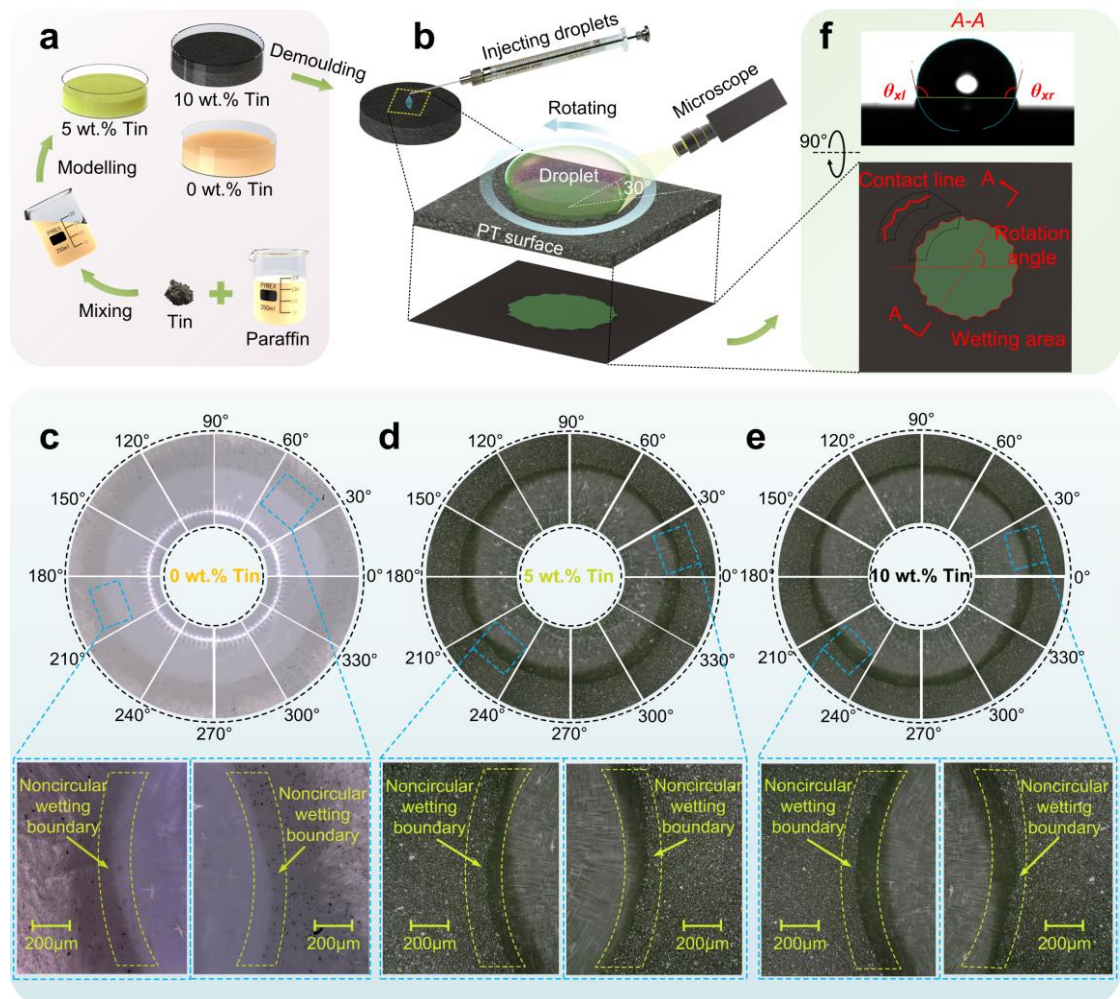

**Figure 1 Noncircular wetting boundary of a droplet on a PT plane. a,** Preparation

of the PT plane. Step 1: Preheat paraffin of 99.9% purity to 100 °C. Step 2: Add tin powder of 99.9% purity (the average particle size is 300 particles per square inch, and a single particle is approximately 48 μm in diameter) to the fused paraffin, mixing them well and then heating to 200 °C. Step 3: Pour the above mixture into a mould (with a surface roughness $S_a$ of 3 μm and a flatness of 5 μm) for modelling. In addition, carry out a series of processes, including precipitation, cooling and ageing at room temperature. Step 4: Demould and obtain three samples with different amounts of tin (Sample 1: PT plane with 0 wt.% tin and a roughness $S_a$ of 0.97; Sample 2: PT plane with 5 wt.% tin and a roughness $S_a$ of 1.53; Sample 3: PT plane with 10 wt.% tin and a roughness $S_a$ of 1.81). **b,** Observation of the noncircular wetting boundary of the droplet on the PT plane. A 5 ml water droplet is dropped on the PT plane, and the plane is rotated 360° at an angle interval of 30°. Meanwhile, a microscope (VHX-5000, Keyence, Osaka, Japan) with a magnification of 1500 is used to photograph the wetting boundary of the droplet. See Extended Data Figures 1–3 for more detail. **c-e,** Observation results of the wetting boundary, with 0° to 360° representing the rotation angles in Figure 1f. **f.** CA measurement principle.

**Variation in the single-point CA along a noncircular wetting boundary**

According to the CA measurement method shown in Figure 1f, the CAs measured along the wetting boundary are shown in Figure 2. The following conclusions are drawn: (a) the CA decreases with increasing hydrophilic substance, (b) the droplet collapses over time[31-36], and (c) there are notable variations in the CA along the wetting boundary. To date, the explanation for the above conclusion (c) is unclear. We suppose that the CA is related to not only the physical properties of the solid and liquid phases at the contact area but also the solid surface morphology. In addition, the external environmental conditions (temperature, pressure, humidity, etc.) also affect the CA. Hence, a single CA at a specific point on the wetting boundary is unable to reflect the actual CAs along the entire wetting boundary. To describe the CA more accurately, the measured CA at a point of the contact line is called the single-point CA ($\theta_{i\text{-}act.}$).

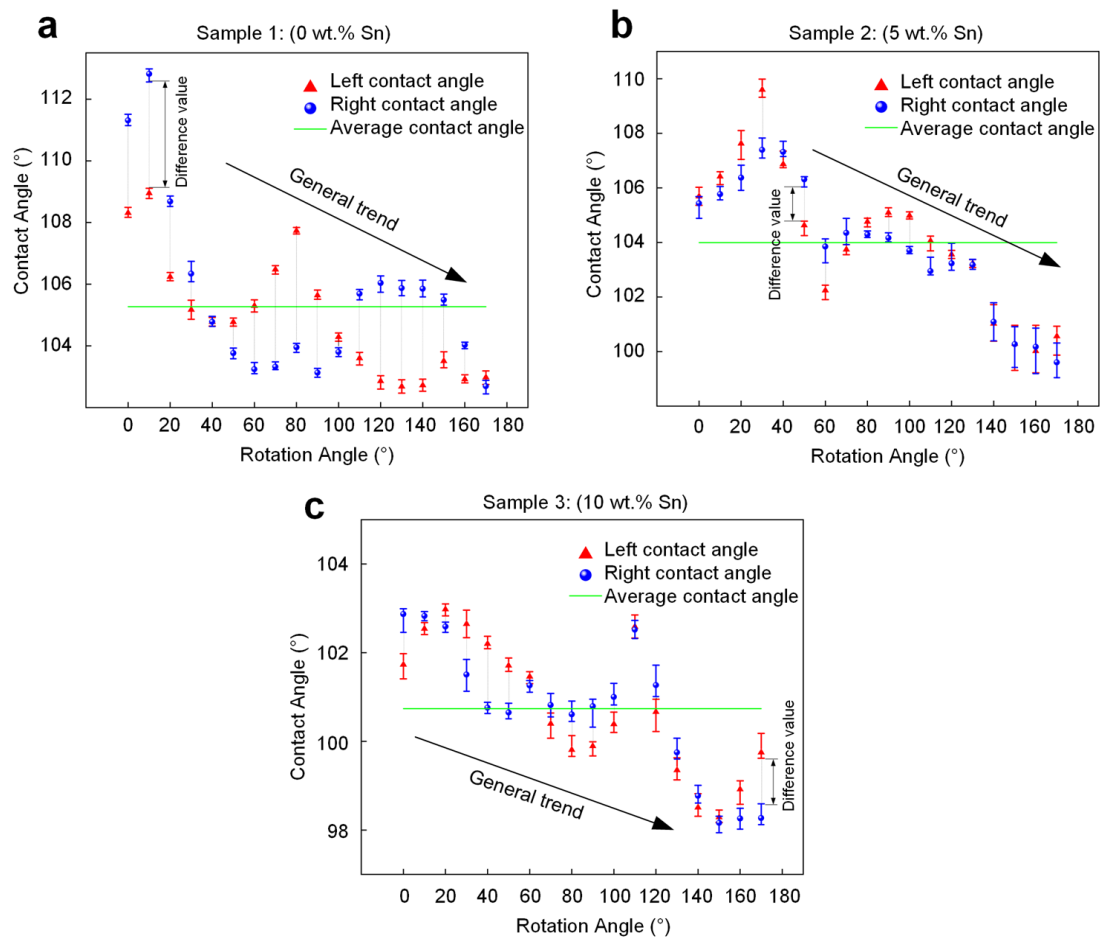

**Figure 2. Measurement results of CAs at different rotation angles.** The experiment is conducted in an indoor environment with a temperature of 22 °C, an atmospheric pressure of 1010 hPa and a relative humidity of 77%. First, the PT plane is placed on a horizontal rotating table, and the deionized water drop (with a volume of 5 μl) is centred

in the plane. The left and right CAs of the droplet are measured by a CA meter (Theta Lite, Biolin, Gothenburg, Sweden). Then, the horizontal rotating table is rotated 360°, and the CAs are measured at intervals of 10°. A total of 36 CAs are obtained, which takes approximately 180 s. At each rotation angle, the left and right CAs are measured 50 times. We take the average of these 50 measurements as the evaluating result in Figure 2. The maximum and minimum CAs obtained from these 50 measurements provide the upper and lower limits of the error bars in the figure, respectively. **a,** On the PT plane with 0 wt.% tin, the measured CAs are distributed from 102.68° to 112.82°, and their mean is 105.27°. **b,** On the PT plane with 5 wt.% tin, the measured CAs range from 99.60° to 109.59°, and their mean is 103.99°. **c,** On the PT plane of 10 wt.% tin, the measured CAs range from 98.16° to 109.61°, and their mean is 100.74°.

To analyse the differences among the single-point CAs, we establish a wetting model of droplets on the ISP and AHSP (see schematic in Figure 3a). On the ISP, the wetting boundary is expected to be an ideal circle, with a single uniform ideal CA ($\theta_{i\text{-}ideal}$) along the circular wetting boundary. On the AHSP, the theoretical model of actual single-point CAs ($\theta_{i\text{-}act.}$) is established on the basis of a noncircular wetting boundary. A noncircular wetting boundary is a combined result of surface energy and surface roughness, which manifests macroscopically as friction acting on the droplet. When spreading on the solid plane and then reaching the equilibrium state, the droplet tends to contract or spread normally due to the line tension $\tau_i$ and is then subjected to the static friction $f$ from the solid surface. We arrive at the following expression for the force on the droplet.

$$f \geq \tau_i \tag{1}$$

We further propose a method to round local actual contact lines into arcs (see the schematic in Figure 3a) to calculate the static friction $f$. A force analysis of the microelement on the wetting boundary is carried out (see schematic in Figure 3c-d). Static friction is related to the external environment (pressure, temperature, height, humidity, etc.) and the vertical component of the gas-liquid interfacial tension $\gamma_{lv}$ (see Formula 2).

$$f = (PL_i - \gamma_{lv} sin\theta_{i-act.} L_i)\eta, \tag{2}$$

where $P$ is the line pressure of the environment and $L_i$ is the arc length of the partially rounded curve. We express the relationship between the arc length $L_i$ and the curvature radius $R_i$ as $L_i = R_i \beta$, where $\eta$ is the friction factor representing the combined apparent result of the surface energy and roughness of the solid surface to obstruct droplet spreading. In addition, Boruvka and Neumann contributed to modifying the Young equation in 1977[7]. They took the line tension $\tau$ into account through differentiation of the three-phase contact line at an angle increment of $\delta\varphi$. The line tension component over the surface tension along the radial direction was derived as $\gamma_\tau = \tau/R_i$, where $R_i$ is the radius of curvature of the local contact line. In the model we built, $R_i$ is defined as the curvature radius of the rounded curve corresponding to the actual contact line. Accordingly,

$$sin\theta_{i-act.} \leq \frac{P}{\gamma_{lv}} - \frac{\tau_i}{\eta \beta R_i \gamma_{lv}} = \frac{P}{\gamma_{lv}} - \frac{\tau_i}{\kappa \gamma_{lv}}, \tag{3}$$

where $\kappa = \eta \beta R_i$ represents the comprehensive influence of local surface properties on the rounded arc of the corresponding local contact line. Importantly, the value of κ is different at various positions on the noncircular wetting boundary. Regardless of gravity, the single-point CA ($\theta_{act.}$) changes with variation in κ at the identical line pressure $P$, gas-liquid surface tension $\gamma_{lv}$ and line tension $\tau_i$. Therefore, the single-point CA varies along the noncircular wetting boundary and fluctuates around the ideal CA ($\theta_{ideal}$). In equation 3, the radius $R_i$ of the rounded curve, as well as the comprehensive influence κ, are directional. Correspondingly, line tension $\tau$ can be positive or negative. The relationship among these variables is shown in Table 1.

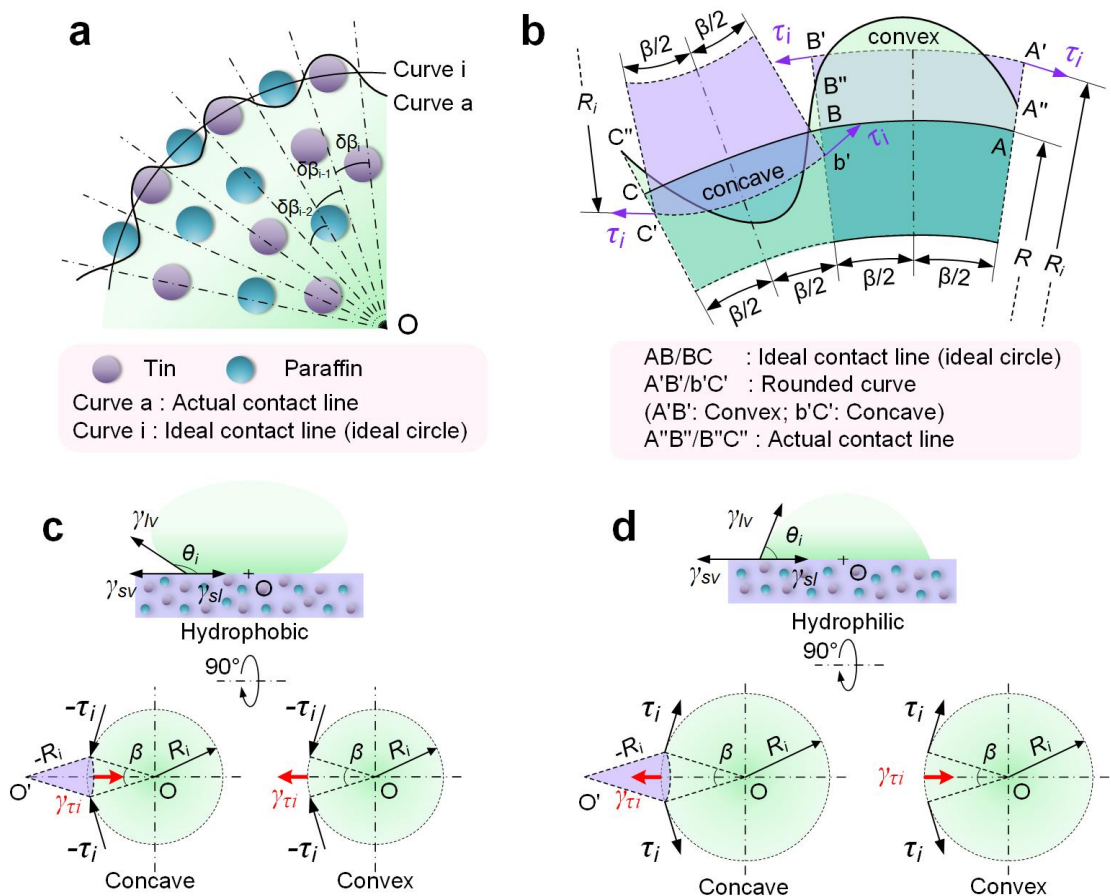

**Figure 3. Influence of the noncircular wetting boundary on the actual CA. a,** Wetting model of a droplet on the ISP and AHSP. **b,** Schematic of the method used to round the actual contact line into an arc. First, the actual wetting area is equivalent to an ideal circle of the same area; in this way, the radius R of the ideal contact line is calculated. Then, taking point O as the origin and the interval angle as β=2π/n, the wetting area is divided into several equal parts through a bisectrix. The bisectrix intersects the ideal contact line and actual contact line at points A, B, C, A", B", and C". The arc lengths of $\widehat{AB}$ and $\widehat{BC}$ are *L*. The lengths of curves A"B" and B"C" are $L_i$. Whether the average curvature of the actual contact curve is greater than that of the ideal curve is then determined. If it is greater, the actual curve is convex (curve A"B"), and through the equivalence principle of area OA"B" to area OA'B', the equivalent radius $R_i$ of the rounded curve (arc A'B') is calculated. If not, it is concave (curve B"C"). Similarly, the area OB'C' is equivalent to the area O'bc, and the radius -$R_i$ ('-' represents that the curve is concave with respect to the geometric centre) of the rounded curve (arc b'C') is also calculated. **c-d,** On the hydrophobic and hydrophilic surface, the line tension is $\tau_i$, the radius of the rounded curve is $R_i$, and the component of line tension $\tau_i$ over the surface tension along the radial direction is $\gamma_{\tau i}$.

**Table 1. Relationship among variable ($\tau_i$, $R_i$, $\gamma_i$, $\sin\theta_i$ and $\theta_i$)**

| $\tau_i$ | Hydrophobic | | Hydrophobic | | Hydrophilic | | Hydrophilic | |
|---|---|---|---|---|---|---|---|---|
| | (-) | | (-) | | (+) | | (+) | |
| $R_i$ | Concave (-) | | Convex (+) | | Concave (-) | | Convex (+) | |
| $\kappa$ | (-) | | (+) | | (-) | | (+) | |
| | ↑ | ↓ | ↑ | ↓ | ↑ | ↓ | ↑ | ↓ |
| $\gamma_i$ | (+) | | (-) | | (-) | | (+) | |
| $\sin\theta_i$ | $0<\theta_i<\pi/2$ | | $\pi/2<\theta_i<\pi$ | | $0<\theta_i<\pi/2$ | | $\pi/2<\theta_i<\pi$ | |
| | ↑ | ↓ | ↓ | ↑ | ↓ | ↑ | ↑ | ↓ |
| $\theta_i$ | ↓ | ↑ | ↑ | ↓ | ↓ | ↑ | ↑ | ↓ |

$\tau_i$, $R_i$ and $\kappa$ are independent variables; $\gamma_i$, $\sin\theta_i$ and $\theta_i$ are dependent variables.

(+/-) represent the direction; (+) represents pointing to the centre of the circle; (-) represents deviating from the centre of the circle.

(↑/↓) represent the variation trend. ↑ represents increase; ↓ represents decrease.

**Evaluation of noncircular wetting**

The ideal CA ($\theta_{ideal}$) cannot be measured in the experiment, and the measured single-point CA ($\theta_{i\text{-}act.}$) reflects the CA at a partial contact line only. In general, researchers previously utilized a single-point CA to reflect the overall wetting condition, which was an oversimplification. To address the previous overgeneralization at wettability, we introduce the concept of average CA ($\theta_{\text{average}} = \frac{\sum_{i=1}^{N}\theta_{i-actual}}{N}$) to more accurately reflect the true comprehensive wettability (see the average CA line in Figure 2). Along the circumferential direction, the CAs of different locations over the contact line are measured at the angle increment of $\delta\beta_i$ ($\delta\beta_i \to 0$), whose average value is defined as the average CA of the wetting boundary. The relationship between the single-point CA and the average CA is similar to the relationship between the instantaneous velocity and the average velocity. The former is discussed from the perspective of the spatial domain, while the latter is discussed in the time domain.

The wettability can be evaluated from two aspects: (a) combining the average CA of the entire wetting boundary and partial wetting boundary or (b) combining the average CA and the fluctuation in the single-point CA. Furthermore, the real complex wetting environment can be reflected through wettability: (a) If the flatness and roughness of the substrate surface are controlled, the properties of the substrate (the composition of the substrate, the distribution of the internal matter particles, etc.) can be investigated, and (b) if the substrate is determined, the properties and states of the droplet and the external environment can be investigated.

**Conclusion and significance**

William Blake's poem goes further[1]: " A heaven in a wild flower." Noncircular boundary could be evaluated by the average CA , which is proved in this study. In summary, we reveal that friction, as a comprehensive influencing factor, reflects the local surface energy and surface roughness and reasonably explains noncircular wetting boundaries. How the single-point CA varying along a noncircular wetting boundary is explored. Ultimately, the average CA, not the single-point CA, is proposed to objectively evaluate wettability.

In the era of Thomas Young, the ideal CA was always approximately substituted by the single-point CA. Although the Cassie, Wenzel and Onda research groups have modified the Young-Dupré equation by considering the actual wetting condition, notable deficiencies remain. Approximately, the single-point CA was utilized to express or reflect the entire wetting condition in their modified models, which was proven to be inaccurate with both theoretical and experimental methods. In the present paper, the ideal CA ($\theta_{ideal}$) and the average CA ($\theta_{aver.}$) are introduced for the first time. Furthermore, the proposed average CA theory unifies the Cassie model (physics) and Wenzel model (engineering), two theories that are currently widely used, eliminating their differences systematically. It is no exaggeration to say that the proposed average CA marks a new milestone for the cognition and measurement of wettability in the field. In terms of methodology, these findings are expected to be conducive to establishing a new evaluation system for CA, which mainly includes the following aspects: (a) setting and control standards for the external environment, (b) standards for measuring single-point CAs, and (c) standards for capturing the contact line.

**References**

1   Young, T. An essay on the cohesion of fluids. Philos. *Trans. R. Soc. Lond.* **95**, 65-87 (1805).

2   Dupré, M. *Théorie mécanique de la chaleur*. (Gauthier Villars, Paris, 1968).

3   Wenzel, R. Resistance of solid surfaces to wetting by water. *Trans. Faraday. Soc.* **28**, 988-994 (1936).

4   Onda, T., Shibuichi, S., Satoh, N. & Tsujii, K. Super-water-repellent fractal surfaces. *Langmuir*. **12**, 2125-2127 (1996).

5   Cassie, A. & Baxter, S. Wettability of porous surfaces. *Trans. Faraday. Soc.* **40**, 546-551 (1944).

6   Gibbs, W. *The Scientific Papers of J. Willard Gibbs. Vol. II: Dynamics, vector*

*analysis and multiple algebra, electromagnetic theory of light*, etc. (dover publications, New York, 1961).

7   Boruvka, L. & Neumann, A. Generalization of the classical theory of capillarity. *J. Chem. Phys.* **66**, 5464-5476 (1977).

8   Chu, Z. & Seeger, S. Superamphiphobic surfaces. *Chem. Soc. Rev.* **43**, 2784-2798 (2014).

9   Bellanger, H., Darmanin, T., Elisabeth, T. & Guittard, F. Chemical and physical pathways for the preparation of superoleophobic surfaces and related wetting theories. *Chem. Rev.* **114**, 2694 (2014).

10  Wang, S., Liu, K. & Jiang, L. Bioinspired surfaces with superwettability: new insight on theory, design, and applications. *Chem Rev.* **115**, 8230-93 (2015).

11  Su, B., Tian, Y. & Jiang, L. Bioinspired interfaces with superwettability: from materials to chemistry. *J. Am. Chem. Soc.* **138**, 1727-48 (2016).

12  Wang, Z., Elimelech, M. & Lin, S. Environmental applications of interfacial materials with special wettability. *Environ. Sci. Technol.* **50**, 2132-2150 (2016).

13  Kung, C., Sow, P., Zahiri, B. & Merida, W. Assessment and interpretation of surface wettability based on sessile droplet contact angle measurement: challenges and opportunities. *Adv. Mater. Interfaces.* **6**, 27 (2019).

14  Wu, W., Zheng, H., Liu, Y., etc. A droplet-based electricity generator with high instantaneous power density. *Nature* **578**, 392-+ (2020).

15  Efzan, M., Ng, W. & Abdullah, M. Effect of fluxes on 60Sn-40Bi solder alloy on copper substrate. *IOP Conf. Ser.: Mate. Sci. Eng.* **133**, 012024 (2016).

16  Lin, Q., Li, F., Jin, P. & Zhong, W. Reactive wetting of TA2 pure Ti and TC4 alloy by molten Al 4043 alloy at 873-973 K. *Vacuum.* **145**, 95-102 (2017).

17  Lu, Y., Li, S., Zuo, W., Ji, Z. & Ding, M. Effect of Cu element addition on the interfacial behavior and mechanical properties of Sn9Zn-1Al(2)O(3) soldering 6061 aluminum alloys: first-principle calculations and experimental research. *J.*

*Alloys Compd.* **765**, 128-139 (2018).

18  Jin, P., Liu, Y., Sun, Q., et al. Wetting of liquid aluminum alloys on pure titanium at 873-973 K. *J. Mater. Res. Technol.* **8**, 5813-5822 (2019).

19  He, S., Gao, R., Li, J., Shen, Y. & Nishikawa, H. In-situ observation of fluxless soldering of Sn-3.0Ag-0.5Cu/Cu under a formic acid atmosphere. *Mater. Chem. Phys.* **239**, 8 (2020).

20  Liu, L., Peng, J., Du, X., et al. Synthesis, composition, morphology, and wettability of electroless Ni-Fe-P coatings with varying microstructures. *Thin Solid Films.* **706**, 7 (2020).

21  Sadeghinezhad, E., Siddiqui, M., Roshan, H. & Regenauer-Lieb, K. On the interpretation of contact angle for geomaterial wettability: contact area versus three-phase contact line. *J. Petrol. Sci. Eng.* **195**, 11 (2020).

22  Xu, Y., Qiu, X., Wang, S., Su, J. & Xing, F. Brazing of tungsten based alloy/superalloy GH907 with lean Cu-Ag based fillers: wetting, interfaces and performance evaluation. *Mater. Lett.* **284**, 5 (2021).

23  Chen, X. & Lu, T. The apparent state of droplets on a rough surface. *Sci. China Phys. Mech.* **52**, 233-238 (2009).

24  Raj, R., Enright, R., Zhu, Y., Adera, S. & Wang, E. Unified model for contact angle hysteresis on heterogeneous and superhydrophobic surfaces. *Langmuir.* **28**, 15777-15788 (2012).

25  Cho, S., Kim, D., Cho, W., Shin, B. & Jeong, M. Diverging effects of topographical continuity on the wettability of a rough surface. *Acs Appl. Mate. Interfaces.* **8**, 29770-29778 (2016).

26  Carmeliet, J., Chen, L., Kang, Q. & Derome, D. Beyond-Cassie mode of wetting and local contact angles of droplets on checkboard-patterned surfaces. *Langmuir.* **33**, 6192-6200 (2017).

27  Pratap, T. & Patra, K. Fabrication of micro-textured surfaces using ball-end micromilling for wettability enhancement of Ti-6Al-4V. *J. Mater. Process. Technol.* **262**, 168-181 (2018).

28  Singh, V., Huang, C., Sheng, Y. & Tsao, H. Smart zwitterionic sulfobetaine silane surfaces with switchable wettability for aqueous/nonaqueous drops. *J. Mater. Chem. A.* **6**, 2279-2288 (2018).

29  Ge, P., Wang, S., Zhang, J. & Yang, B. Micro-/nanostructures meet anisotropic wetting: from preparation methods to applications. *Mater. Horizons.* **7**, 2566-2595 (2020).

30  Xu, C., Jia, Z. & Lian, X. Wetting and adhesion energy of droplets on wettability gradient surfaces. *J. Mater. Sci.* **55**, 8185-8198 (2020).

31  Kong, J., Yung, K., Xu, Y. & Tian, W. Wettability transition of plasma-treated polystyrene micro/nano pillars-aligned patterns. *Express Polym. Lett.* **4**, 753-762 (2010).

32  Hsu, C., Su, T., Wu, C., Kuo, L. & Chen, P. Influence of surface temperature and wettability on droplet evaporation. *Appl. Phys. Lett. 106*, 5 (2015).

33  Long, J., Zhong, M., Zhang, H. & Fan, P. Superhydrophilicity to superhydrophobicity transition of picosecond laser microstructured aluminum in ambient air. *J. Colloid Interface Sci.* **441**, 1-9 (2015).

34  Pan, Y., Kong, W., Bhushan, B. & Zhao, X. Rapid, ultraviolet-induced, reversibly switchable wettability of superhydrophobic/superhydrophilic surfaces. *Beilstein J. Nanotechnol.* **10**, 866-873 (2019).

35  Wang, T., Si, Y., Luo, S., Dong, Z. & Jiang, L. Wettability manipulation of overflow behavior via vesicle surfactant for water-proof surface cleaning. *Mater. Horizons* **6**, 294-301 (2019).

36  Cheng, H., Chang, T., Lin, C. & Chen, P. Wettability of laser-textured copper surface after a water-bath process. *Aip Adv.* **9**, 7 (2019).

**Data availability**

The datasets generated or analyzed during the current study are available from the corresponding author on reasonable request.

**Code availability**

The code generated during the current study is available from the corresponding author on reasonable request.

**Author contributions**

Jianhui Zhang discovered the phenomenon, proposed and designed this research. Jianhui Zhang, Xiaosheng Chen, Zhenzhen Gui and Zhenlin Chen contributed to the experimental method and experimental design as well as data analysis. Mingdong Ma, Yuxuan Huo, Weirong Zhang and Fan Zhang carried out the experiment and collected the experimental data. Xiaosi Zhou and Qian Huang conducted repeating experiments and data processing. Jianhui Zhang and Xiaosheng Chen conducted theoretical analysis. Jianhui Zhang, Xiaosheng Chen and Zhenzhen Gui prepared the full paper. All authors analyzed the results and discussed the manuscript.

**Competing interests**

The authors declare no competing interests.

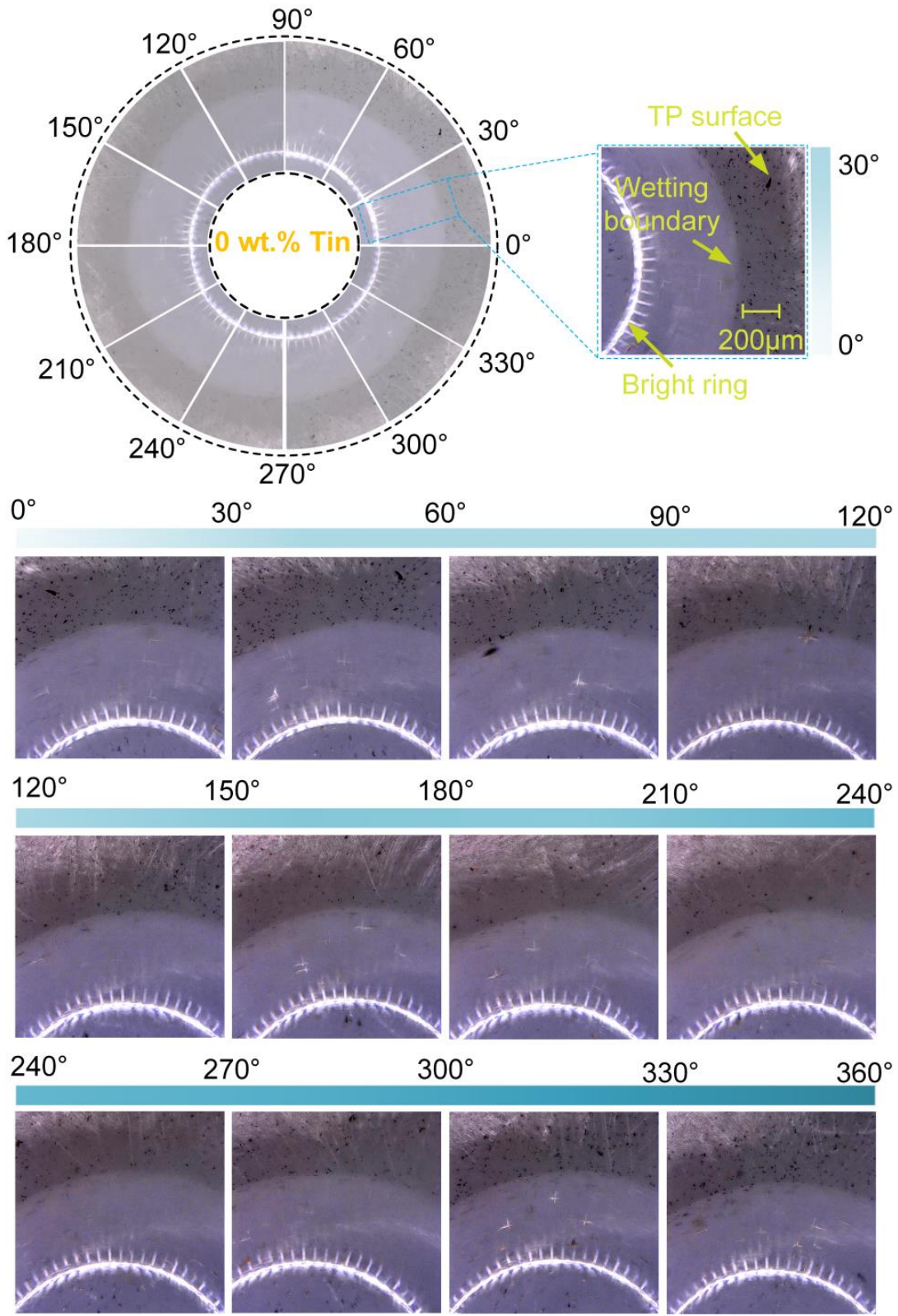

**Extended Date Figure 1 Wetting boundary in PT plane with 0 wt.% tin.**

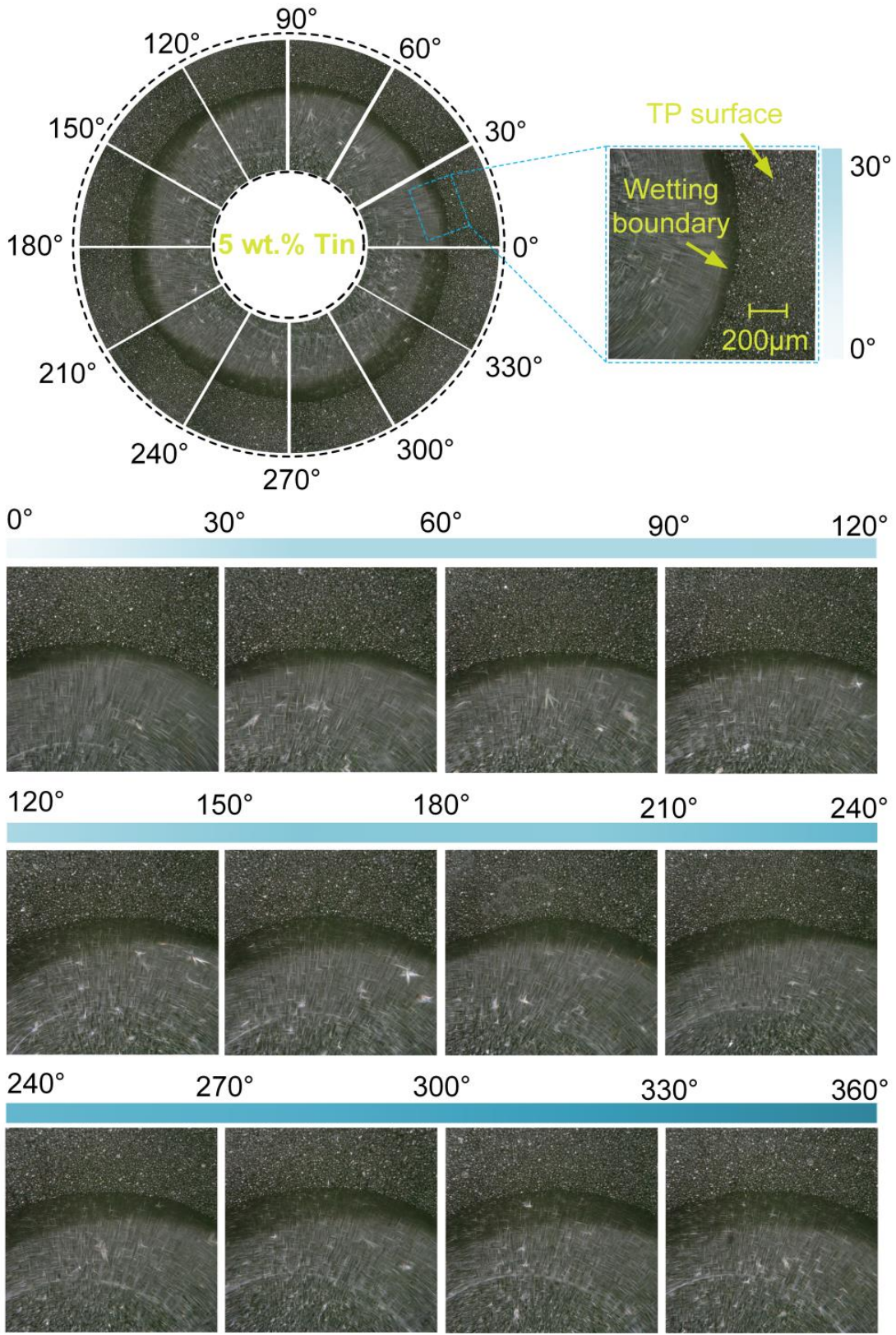

**Extended Date Figure 2 Wetting boundary in PT plane with 5 wt.% tin.**

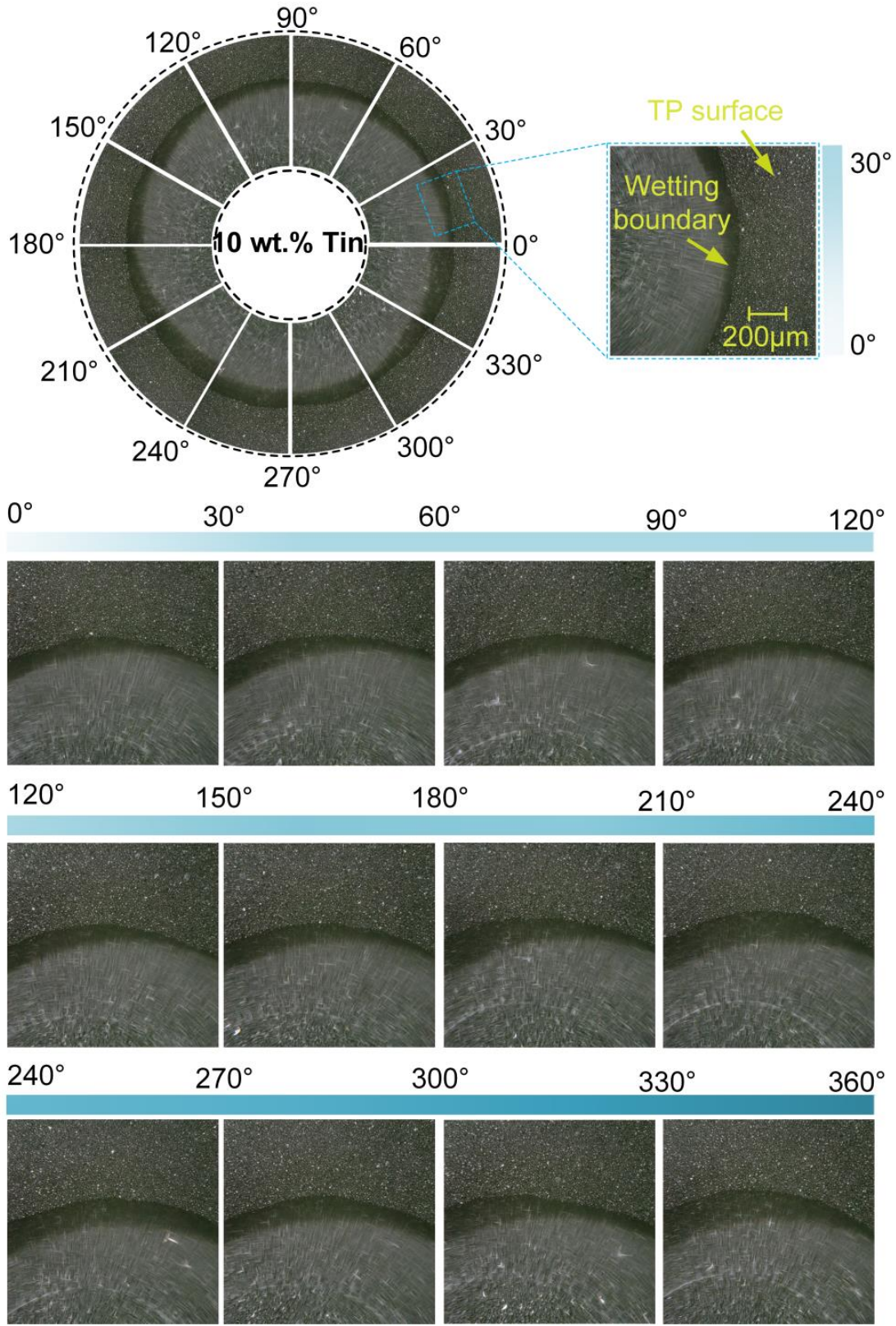

**Extended Date Figure 3 Wetting boundary in PT plane with 10 wt.% tin.**